\documentclass[sigconf, screen]{acmart}

\usepackage{graphicx}
\usepackage{textcomp}
\usepackage{xcolor}
\usepackage{url}
\usepackage{multirow}

\usepackage{balance}
\usepackage{tikz}
\usepackage{caption}
\usepackage{subcaption}
\usepackage{booktabs}
\usepackage{svg}
\usepackage[colorinlistoftodos]{todonotes}
\usepackage[linesnumbered,ruled,vlined]{algorithm2e}

\AtBeginDocument{%
  \providecommand\BibTeX{{%
    \normalfont B\kern-0.5em{\scshape i\kern-0.25em b}\kern-0.8em\TeX}}}

\setcopyright{acmlicensed}
\acmPrice{15.00}
\acmDOI{10.1145/3611643.3616316}
\acmYear{2023}
\copyrightyear{2023}
\acmSubmissionID{fse23main-p673-p}
\acmISBN{979-8-4007-0327-0/23/12}
\acmConference[ESEC/FSE '23]{Proceedings of the 31st ACM Joint European Software Engineering Conference and Symposium on the Foundations of Software Engineering}{December 3--9, 2023}{San Francisco, CA, USA}
\acmBooktitle{Proceedings of the 31st ACM Joint European Software Engineering Conference and Symposium on the Foundations of Software Engineering (ESEC/FSE '23), December 3--9, 2023, San Francisco, CA, USA}
\received{2023-02-02}
\received[accepted]{2023-07-27}

\begin{document}
\newcommand{\name}{\texttt{Outage-Watch}}
\newcommand{\tobedone}[1]{\textcolor{blue}{#1\\}}
\newcommand\mycommfont[1]{\footnotesize\ttfamily\textcolor{blue}{#1}}

\newcommand{\shiv}[1]{\textcolor{red}{#1\\}}
\newcommand{\sarthak}[1]{\textcolor{blue}{#1\\}}
\newcommand{\shubham}[1]{\textcolor{orange}{#1\\}}
\newcommand{\ourmethod}{\texttt{Outage-Watch}}
\newcommand{\etal}{\textit{et al.}}

\newcommand{\indep}{\perp \!\!\! \perp}

\newcommand*\circled[1]{\tikz[baseline=(char.base)]{
            \node[shape=circle,fill={rgb:black,2;white,3},inner sep=1pt] (char) {
            % \footnotesize \bf{ 
            \textcolor{white}{#1}
            % }
            } ;}}
            
\title{\ourmethod{}: Early Prediction of Outages using Extreme Event Regularizer}


\author{Shubham Agarwal}
\affiliation{%
  \institution{Adobe Research}
  \city{Bangalore}
  \country{India}}
\email{shagarw@adobe.com}

\author{Sarthak Chakraborty}
\authornote{Work done at Adobe Research, India}
\affiliation{%
  \institution{University of Illinois Urbana-Champaign}
  \city{Champaign}
  \country{USA}}
\email{sc134@illinois.edu}

\author{Shaddy Garg}
\affiliation{%
  \institution{Adobe}
  \city{Bangalore}
  \country{India}}
\email{shadgarg@adobe.com}

\author{Sumit Bisht}
\authornotemark[1]
\affiliation{%
  \institution{Amazon}
  \city{Bangalore}
  \country{India}}
\email{bishts002@gmail.com}

\author{Chahat Jain}
\authornotemark[1]
\affiliation{%
  \institution{Traceable.ai}
  \city{Bangalore}
  \country{India}}
\email{chahatjain99@gmail.com}

\author{Ashritha Gonuguntla}
\authornotemark[1]
\affiliation{%
  \institution{Cisco}
  \city{Bangalore}
  \country{India}}
\email{ashrithag.0907@gmail.com}

\author{Shiv Saini}
\affiliation{%
  \institution{Adobe Research}
  \city{Bangalore}
  \country{India}}
\email{shsaini@adobe.com}

\renewcommand{\shortauthors}{S Agarwal, S Chakraborty, S Garg, S Bisht, C Jain, A Gonuguntla, S Saini}

\begin{abstract}

Cloud services are omnipresent and critical cloud service failure is a fact of life.  In order to retain customers and prevent revenue loss, it is important to provide high reliability guarantees for these services. One way to do this is by predicting outages in advance, which can help in reducing the severity as well as time to recovery. It is difficult to forecast critical failures due to the rarity of these events. Moreover, critical failures are ill-defined in terms of observable data. Our proposed method, \ourmethod{}, defines critical service outages as deteriorations in the Quality of Service (QoS) captured by a set of metrics. \ourmethod{} detects such outages in advance by using current system state to predict whether the QoS metrics will cross a threshold and initiate an extreme event. A mixture of Gaussian is used to model the distribution of the QoS metrics for flexibility and an extreme event regularizer helps in improving learning in tail of the distribution. An outage is predicted if the probability of any one of the QoS metrics crossing threshold changes significantly. Our evaluation on a real-world SaaS company dataset shows that \ourmethod{} significantly outperforms traditional methods with an average AUC of $0.98$. Additionally, \ourmethod{} detects all the outages exhibiting a change in service metrics and reduces the Mean Time To Detection (MTTD) of outages by up to $88\%$ when deployed in an enterprise cloud-service system, demonstrating efficacy of our proposed method.
\end{abstract}



\maketitle

%-------------------------------------------------------------------------------
\section{Introduction} \label{sec:introduction}

The use of cloud services for deploying applications has seen a tremendous growth. According to a recent report~\cite{cloud-adoption}, about 94\% of enterprises already use cloud services.
% The shift to cloud computing has resulted in the rise of software systems being hosted on cloud platforms. \shiv{USE THIS INSTEAD OF PREVIOUS SENTENCE: According to a recent report, about 94\% of enterprises already use cloud services.} \sarthak{What is the citation of the report?} 
However, these cloud services, with numerous components, are complex and prone to failures~\cite{cloud_failures,cloud_failures_big} and outages~\cite{cloud_outages,cloud_outages_2} due to frequent updates, changes in operation, repairs, and device mobility. 
% Cloud providers offer services with specific Quality of Service (QoS) requirements and any failure to meet these standards results in Service Level Agreements (SLA) violations \cite{Google_SRE}, revenue loss and customer dissatisfaction.
Cloud providers offer services with specific Quality of Service (QoS) requirements, which are technical specifications defining various aspects of system quality, such as performance, availability, scalability, and serviceability. These QoS requirements are driven by business needs outlined in the business requirements. Any failure to meet these predefined QoS standards can lead to Service Level Agreement (SLA) violations, resulting in revenue loss and customer dissatisfaction \cite{Google_SRE}.
A study\footnote{\url{https://www.lloyds.com/news-and-insights/risk-reports/library/cloud-down}} found that a 3-6 day outage by a leading cloud provider in the US could result in \$15 billion loss. As a result, cloud system reliability is critical for business success, as outages can severely impact QoS metrics (resource availability, latency, etc.) resulting in compromised system availability and a poor user experience.

Several monitoring and alerting tools (refer \S \ref{sec: Background and Problem Definition}) are employed to monitor and ensure the performance of cloud services. Automating system troubleshooting has been found to improve reliability, efficiency, and agility for enterprises \cite{chen2020towards,dang2019aiops,lou2013software}. Despite these efforts, cloud systems still experience incidents and outages \cite{Amazon,Azure, Google}. Timely detection and remediation of outages is essential for reducing system downtime. However, a reactive approach to incident detection is often used in practice, hindering effective outage management~\cite{li2021fighting, airalert, eWarn}. With a possible innovation in being able to predict the outages well in advance, the time to detect these outages can be reduced significantly.

\begin{figure}[t]
    \centering
    \includegraphics[width=0.4\textwidth]{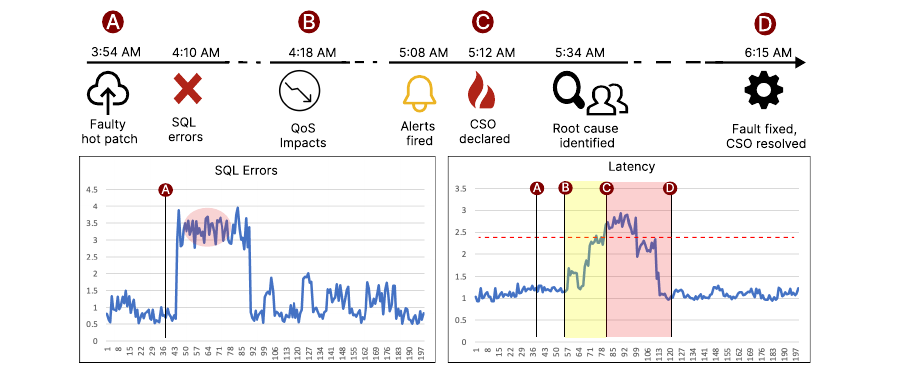}
    \caption{Illustration of the life-cycle of an outage. A refers to the point when the root cause of a fault occurred, B represents the time when it started affecting the performance metrics. When the metrics crossed their respective thresholds, alerts fired which led to an outage being declared at C. The time between C to D is when the engineers diagnose and resolve the issue. The plots below show the variation in the root cause metric and the QoS metric at these times.}
    \label{fig:introduction-figure}
\end{figure}

Consider a real-world scenario in Figure~\ref{fig:introduction-figure} showing the timeline of an outage caused by a flawed configuration change in a Storage service.
In this scenario, a 3:54 am (A) failure sparked a sequence of problems, including SQL errors at 4:10 am and an increase in latency that starts affecting the QoS at 4:18 am (B). Alerts were triggered at 5:08 am when latency exceeded pre-defined thresholds. It took nearly 55 minutes (from 4:18 am (B)) to realize it was a cross-service issue and declare an outage at 5:12 am (C). An experienced Site Reliability Engineer (SRE) \cite{Google_SRE, Atlassian} was engaged to mitigate the issue, which was resolved at 6:15 am (D) with all services back to normal. Here, the flawed change impacted several SQL databases and spread to other services. The current reactive approach relying on alerts showed significant delay in detecting the outage, as seen by the ramp up in underlying metrics affecting QoS between 4:18 am (B) and 5:12 am (C).
This example highlights the potential to predict a substantial fraction of outages in advance by utilizing the information available during the ramp-up phase.
 % \textcolor{red}{In light of the limited downtime budget of only 500 to 50 minutes in a year, corresponding to up-time guarantees of 99.9\% and 99.99\%, detecting an outages hours or even minutes in advances can provide large dividends.} In this paper, our goal is to provide an end-to-end approach for reducing the mean time to detection (MTTD) by detecting outages early.
In consideration of the strict downtime constraints, with only 500 to 50 minutes of allowable downtime per year corresponding to the uptime guarantees of 99.9\% and 99.99\% respectively, the early detection of outages, even minutes in advance, can result in significant benefits. The objective of this paper is to present a comprehensive solution aimed at reducing the mean time to detection (MTTD) through early detection of outages.

Outages manifest in two major ways, (i) as degradation in QoS and other metrics, (ii) detected only through user reports and do not manifest in observable metrics. The first type of outages, accounting for 50-70\% of the incidents as observed from our data (see \S\ref{sec:implement}), exhibit characteristics that allow for prediction. However, outage prediction in cloud systems is a complex task due to the vast number of interdependent metrics. An SRE, who traditionally detect outages using rule-based alerts, often only have a limited view of the overall system, leading to difficulties in quickly and accurately identifying issues. Such approaches rely on human knowledge and is insufficient for large-scale production cloud systems, which have a vast number of complex and ever-changing rules. Our interviews with engineers from various service teams revealed that detection could take hours, particularly in cases where there are multiple concurrent alerts. This highlights the importance of developing a more efficient method for predicting outages in cloud systems.

Previous works \cite{airalert, li2021fighting, ikram2022root} on failure prediction through runtime monitoring which requires a substantial amount of data from the faulty state of the system are not applicable in this scenario, as outages are rare events \cite{Justamom56} and hence, data is not available in the faulty state in plenty. In addition, using alerts to detect outages takes a toll on MTTD since they are fired after a significant ramp-up in metrics has been identified. Failure detection literature from other domains~\cite{anantharaman2018large, lu2020making} are not extensible to our case since the nature and the quantity of failures is very different in an enterprise service. Our scenario has very few outages and directly extending those works fail.

In this work, we propose a novel system (\ourmethod{}) for predicting outages in cloud services to enhance early detection. We define outages as extreme events where deterioration in the QoS captured by a set of metrics goes beyond control. \ourmethod{} models the variations of QoS metrics as a mixture of Gaussian to predict their distribution. We also introduce a classifier that is trained in a multi-task setting with extreme value loss to learn the distribution better at the tail, thus acting as a regularizer~\cite{scholkopf2002learning}. \ourmethod{} predicts an outage if there is a significant change in the probability of the QoS metrics exceeding the threshold. Our evaluation on real-world data from an enterprise system shows significant improvement over traditional methods with an average AUC of 0.98. Furthermore, we deployed \ourmethod{} in a cloud system to predict outages, which resulted in a 100\% recall and reduced MTTD by up to 88\% (20 - 60 minutes reduction).

Our major contributions can be summarized as follows:
\begin{enumerate}
    
    \item We propose a novel approach \ourmethod{} to predict outages in advance, which are manifested as large deteriorations in a chosen set of metrics (QoS) reflecting customer experience degradation. \ourmethod{} works even in the absence of actual outages in training data.
    
    \item \ourmethod{} generates the probability of a metric crossing any threshold, making it flexible to define the threshold, unlike classification tasks. It predicts the distribution of QoS metric values in future given current system state, and improves learning the tail distribution via extreme value loss to capture outages before they happen.
   
    \item An evaluation of the approach on real service data shows an improvement of $7-15\%$ over the baselines in terms of AUC, while its deployment in a real setting was able to predict all the outages which exhibited any change in the observable metrics, thus reducing the MTTD.
    
\end{enumerate}

The rest of the paper is organized as follows. We briefly talk about related works in Section \ref{sec:related-work} followed by the background and problem formulation in Section \ref{sec: Background and Problem Definition}. In Section \ref{sec:approach_motivation} and \ref{sec:approach} we outline the motivation  and describe \ourmethod{}. With Section \ref{sec:results} analyzing its performance, we conclude in Section \ref{sec:conclusion}.

% \begin{enumerate}
%     \item We study the problem of outage prediction and identify the obstacles that result in prolonged outage detection time and the potential opportunities to predict outages early .
%     \item We propose \name{} to predict outages by using current system state to predict whether the QoS metrics will cross a threshold.
%     \item  We evaluate \name{} with real-world data collected. The evaluation results confirm the effectiveness of the proposed approach by using test deployed \ourmethod{}.
    
%     \shubham{@sarthak, just drafted as a skeleton, need to see this part and add to contributions}
% \end{enumerate}

% application contributions to be mentioned as well

    % \item \textbf{Write about the technical challenges in doing the same and show that others related paper have not solved the problem} - \textbf{label issues} - supervised learning - outages are different.

% As illustrated in Figure \ref{fig:introduction-figure}, a fault is labeled as an outage only after it has escalated to the point where the customer experience begins to deteriorate or the customer reports an issue. However, such reactive approaches of labeling outages often increases the mean time to detect those outages. By detecting outages in advance, diagnosis and resolution can be initiated in a timely manner. This highlights the need for a proactive system that reduces the mean time to detect outages.

%-------------------------------------------------------------------------------

%-------------------------------------------------------------------------------
\section{Related Work} \label{sec:related-work}

Service reliability has been a well-researched area in both academia and industry~\cite{airalert, li2021fighting, ikram2022root, chakraborty2023esro, causeinfer}. Several works have attempted to address the problems of detecting, localizing, and mitigating outages and failures.
Alerts are often used in detecting outages~\cite{lin2014unveiling,jiang2009ranking,zhao2020automatically}, which are triggered when a system fault occurs and metric values crosses a threshold.
Recent approaches such as AirAlert~\cite{airalert}, Fog of War~\cite{li2021fighting}, and eWarn~\cite{eWarn} compute features based on alerts and predict outages using tree-based models. However, alerts only trigger when a system is already in a critical state, thus incurring low reduction in MTTD.

Previous research in the area of time series forecasting constitutes a relevant body of literature since changes in metric value time series can be forecasted to detect outages. Classical auto-regressive models~\cite{gurland1954hypothesis, arima} to predict future metric values have limitations that have been overcome by recent advancements in deep neural network (DNN) based models, such as RNNs, LSTMs and GRUs, which have proven to be more effective in modelling time series data~\cite{siami_arima, dasgupta2017nonlinear, yang2015deep, lin1996learning, hochreiter1997long, cho2014learning}.
Empirical evidence supports the use of these deep recurrent models for time series prediction \cite{lin2017hybrid, qin2017dual, zhu2017next}.
However, they perform poorly~\cite{laptev2017time} in predicting rare events like outages due to imbalanced data \cite{laptev2017time, lin2017focal}, also known as extreme events \cite{albeverio2006extreme, ghil2011extreme}. Predicting extreme events remains a challenging and active area of research \cite{friederichs2012forecast, haan2006extreme}.

Recent studies~\cite{laptev2017time, ding2019modeling} have attempted to address the challenge of forecasting extreme events in time series data through innovative DNN architectures. \cite{laptev2017time} employs an auto-encoder for feature extraction while \cite{ding2019modeling} uses an Extreme Value Loss (EVL) function. However, modelling point prediction of extreme events as classification task would need re-estimation of the model if the threshold definition changes for defining such events. However, \name{} sets itself apart by combining the EVL function with LSTM and predicting the future distribution of events using a mixture normal network, thus allowing a flexible change of threshold.

In the domain of reliability engineering, prior studies have concentrated on detecting anomalies in time series data for monitoring systems \cite{chen2022adaptive}. 
Supervised anomaly detection models like \cite{chen2019outage, park2019fault, li2021fighting} are effective in predicting anomalies, but they require a substantial amount of labelled data, making them unsuitable for our application. On the other hand, unsupervised methods~\cite{zhang2019cross} can be used to identify anomalies in data without the need for labelled data. More recently, ML-based models including autoencoders and transformers are used for anomaly detection in seasonal metric time series~\cite{tuli2022tranad, su2019robust,park2018multimodal}. However, they are limited to detecting events as they happen and lack interpretability to distinguish significant service degradation. Our approach differs by predicting the probability of metric exceeding threshold through its probability distribution analysis, instead of simply detecting whether the actual metric crosses a threshold or not, as in traditional anomaly detection.

In related literature of failure detection, several efforts have been made to predict specific types of failures, utilizing large sets of system logs and metrics~\cite{zhang2018prefix, oki2018mobile, xu2018improving}. Our approach differs in predicting general incidents based on a limited set of relevant metrics, determined by domain knowledge. Previous works like \cite{anantharaman2018large, dos2017predicting,lu2020making} have used ML and DNN techniques to predict disk failures. A comparison of disk failure prediction methods was presented in \cite{lu2020making}. However, they are not suitable for predicting rare and extreme outages. Our approach, based on distribution forecasting, is flexible and outperforms some of the baselines discussed in \cite{lu2020making}.

% \begin{enumerate}

% \textbf{Time series forecasting and extreme event forecasting}
%     \item\textbf{Outage Detection using alerts:} AirAlert, eWarn, Fog of War. How our approach is different as we do not work to detect issues based on alerts fired data and how metrics data can be used to predict earlier. 
%     \item\textbf{Anomaly Detection on Metrics Data:} Refer to supervised and unsupervised anomaly detection for metrics data. How our approach is different as labelled data is not available.Also, Anomaly detection is like alerts on univariate time series and is not applicable for extreme event prediction as not all of them can be signalled as potential outages. Also, it is on-time detection not prediction

%     \item\textbf{Failure detection:} Disk failure literature and why they are not directly applicable.
%     \item\textbf{Change based incident detection}: Outages due to bad change roll-outs, they happen immediately and we are not working to solve them.
% \end{enumerate}
%-------------------------------------------------------------------------------

%-------------------------------------------------------------------------------
\vspace{-0.2cm}
\section{Background and Problem Definition}
\label{sec: Background and Problem Definition}
In this section, we briefly describe basic monitoring system concepts, followed by the problem formulation.

\textbf{Monitoring Metrics:} In any enterprise level service deployment, reliability engineers monitor the performance of these deployed services. Various monitoring tools like grafana~\cite{grafana},  new-relic~\cite{newrelic}, splunk~\cite{splunk}, etc. are employed to monitor the services and collect service metrics that can be used to measure the performance of the system. Monitoring metrics provide the most granular information, with these tools recording them at specific time intervals.

\textbf{Alerts:} Alerts are defined on monitoring metrics. Whenever these metrics cross a certain pre-defined threshold defined by the reliability engineers, an alert is triggered. They represent system events that require attention, such as API timeouts, exceeding response latencies, service errors and network jitters. An alert contains several fields like the alert definition, time of alert, which service fired the alert, text description of the alert, datacenter region where the alert was fired and severity level which can be "high", "medium" or "low".

\textbf{Extreme Events: }An extreme event is characterized by a set of monitoring metrics exceeding their respective thresholds (95\%-ile metric threshold based on our data), causing unusual system behaviour. This results in multiple alerts with varying levels of severity being triggered, prompting engineers to take action. It's important to note that not all extreme events lead to outages, as they may not necessarily indicate a fault in the system. Extreme events which lead to outages cause violations in Service Level Agreements (SLAs). Extreme events can be perceived as situations when the system starts showing signs of degradation, and the metric values surpass their 95 or 99\%-ile thresholds. In addition, it is important to note that extreme events are not recorded as separate incidents. We generate proxy labels (see \S\ref{sec:label-generate}) to identify the extreme events.

\textbf{Outage:} In general, outages are declared under severe situations, when extreme events persist for a long time, leading to a degradation in the service quality and often leading to a customer impact. Often, outages affect multiple services where a fault propagates based on service dependency. Mitigating outages require cross-team collaboration. An outage usually triggers correlated alerts~\cite{chen2020identifying, zhao2020understanding} and escalates from a single or a group of alerts. Not all extreme events manifest as an outage since engineers manage these systems through constant monitoring, and they identify and intervene in the system during potential issues even before they escalate into full-scale outage.

\subsection{Problem Formulation}
We now formally define our problem statement. Metrics $\mathcal{M}_{tot}$ are continuously monitored in the system, essentially forming a time series. The task is to understand the trends in some or all of the metrics in $\mathcal{M}_{tot}$ and predict an impending outage, with the goal of predicting it as early as possible. It is obvious that a change in the metrics will show up only when a fault has occurred. Thus, the goal of an outage forecasting solution is to minimize the lag time between the actual occurrence of the fault and its identification as an extreme event, while also minimizing false positive cases.

With $t$ as the wall-clock time, the input to the outage forecasting module is a set of relevant metrics (see \S\ref{sec:qos_metrics}) from $[t-w, t]$ where $w$ is the window length. More details on the pre-processing of metrics is elucidated in \S\ref{sec:metric_preprocess}. With ground-truth labels generated based on the occurrence of extreme events, the supervised ML model \ourmethod{} aims to forecast an outage by learning the distribution of the relevant metrics at a certain time in the future. A distribution of metrics is essentially a probability density function of the metric values at a certain time.
%-------------------------------------------------------------------------------

%-------------------------------------------------------------------------------
\section{Solution Motivation} \label{sec:approach_motivation}

In this section, we present the rationale behind our solution design and provide a concise overview of how it functions.

\begin{figure}[h]
    \centering
    \includegraphics[width=0.5\textwidth]{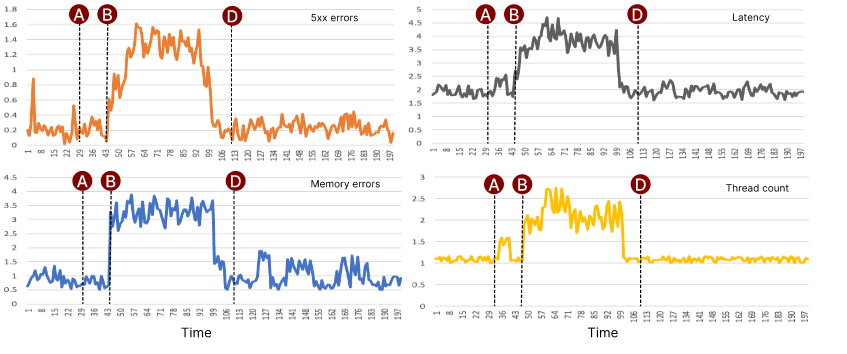}
    \caption{Multiple performance metrics, some of which are QoS metric gets impacted during an outage. The collective information from all these metrics help in detecting the outage. A, B and D corresponds to the time defined in Fig. \ref{fig:introduction-figure}}
    \label{fig:metric_variations}
    \vspace{-0.52cm}
\end{figure}

\begin{figure*}[t]
    \centering
    \includegraphics[width=1.0\linewidth]{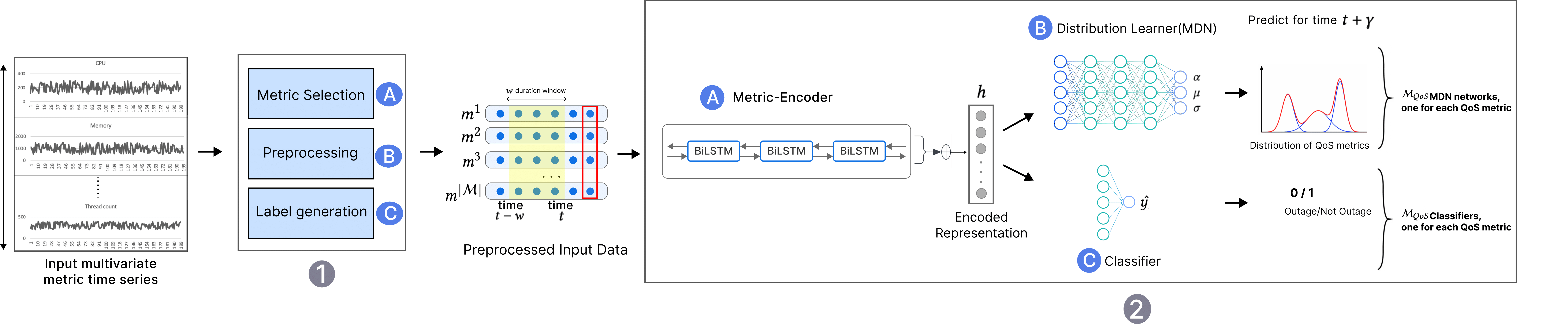}
    \caption{Overall Architecture of \ourmethod{}, comprising of (1) metric processing phase and the (2) distribution learning phase. The distribution and label prediction generated from (2) at time step $t$ are evaluated against ground-truth metric value and labels from a future time step $t+\gamma$, which we get from (1B) and (1C) respectively.}
    \label{fig:overview_fig}
\end{figure*}

\subsection{Design Motivation} \label{sec:design-motiv}
As discussed in \S\ref{sec:introduction}, outage prediction models should aim to predict the probability of an outage in advance to ensure timely recovery during a fault. One way to achieve this is to monitor the system metrics for deviations from their regular trend, as these deviations are often indicative of an outage. However, the current system monitoring tools often fail to detect deviations until they surpass a specific threshold and activate an alert, provided that an alert has been defined for those system metrics. However, proactive monitoring of system metrics allows for earlier and more efficient identification of outages. This motivates the design of our proposed approach, \ourmethod{}.

We have observed from the data (Figure \ref{fig:metric_variations}) that during an outage, multiple metrics \cite{AnatomyO31} show disturbances and deviations from trends simultaneously. These metrics progressively become more extreme and affect multiple interdependent services. \ourmethod{} takes multiple monitoring metrics as input and aims to predict the future behaviour or change in the values of certain metrics (which we refer to as Quality of Service (QoS) metrics, and will be further defined in \S\ref{sec:qos_metrics}). By predicting the distribution of these metrics in advance, we can compute the probability of an upcoming outage based on the likelihood of the metric value crossing a certain threshold. These thresholds can be dynamic and may be defined by Service Level Agreements (SLAs) on QoS metrics for each service. 

\textbf{Why do we need to learn the distribution of metrics?} Outages are often scarce in any established production-level service. Hence, if the outage prediction task is modelled as a simple classification problem, a machine learning model will tend to fail even after the imbalance in the dataset is addressed, because of the extremely skewed distribution of data points coming from one class (outage occurrences). For example, it is often the case that one observes only two outages over a period of 6 months. However, one way to circumvent this issue is to design our problem as a distribution learning task. We then have a corresponding metric value at every timestamp of the data, and a learned regression model can predict a metric value at any other time stamp.  We can also use the same strategy to learn a more accurate distribution in the tail, where extreme events are often manifested.
This allows the flexibility of constructing the entire distribution and using variable threshold based on Service Level Agreements (SLA) requirements.

We provide further technical details on how the model predicts the distribution of the relevant metrics and how it can be used to predict outages in subsequent sections.

\subsection{Architecture Overview} \label{sec:overview}
% \todo[inline]{Overall architecture, several components, how will it look when it is deployed, what components constitute the training and inference, better if we can label some big modules and then explain each of the modules in the following section}

The proposed framework of \ourmethod{} (Fig. \ref{fig:overview_fig}) comprises two main phases: a metric processing phase (denoted as \circled{1} in Fig. \ref{fig:overview_fig} and a distribution learning phase (denoted as \circled{2} in Fig. \ref{fig:overview_fig}). Module \circled{1} first selects the relevant metrics that will be used to forecast the outages, pre-processes them and then generates \textit{proxy labels}. What we mean by \textit{labels} here and why do we need to generate them will be elucidated in details in \S\ref{sec:label-generate}. On the other hand, module \circled{2} forecasts the outage by predicting the distribution of the relevant metrics selected from phase \circled{1} at a future time.

In order for the distribution learner to predict outages at a future time $t+\gamma$ where $\gamma$ is the prediction look-ahead, the machine learning algorithm must learn from the appropriate ground truth. Thus, with input data at $t$, the ground truth is constructed such that our method can predict the distribution of the metrics  at time $t+\gamma$ and hence predict potential outages. The distribution learner framework first uses a metric encoder to encode the system state in the past trend of monitoring metrics, and then the distribution of the relevant metrics is forecasted using a Mixture Density Network (MDN). MDN aims to predict the parameters of the forecasted distribution. Additionally, the framework uses a classifier as an extreme value regularizer for better learning in the tail of the distribution. The technical details are presented in \S\ref{sec:dist-forecasting}.

Though the pipeline is trained to predict the parametric distribution of metrics, inference on whether and when the outage is being detected need to be developed. \S\ref{sec:inference} focuses on identifying the likelihood of outages by evaluating the probability of each of the relevant metric crossing a defined threshold. If there is a sudden increase in the probability of exhibiting extreme values based on the distribution learnt, \ourmethod{} takes this as an indication of an outage. A thresholding mechanism is employed on the probability value to predict an outage.

\section{\ourmethod{}} \label{sec:approach}

In \S\ref{sec:overview}, we have discussed the overall architecture of \ourmethod{} and talked about its two main components briefly.  We shall now delve into the details of each component.

\subsection{Metric Processing}

\subsubsection{Metric Selection and Quality of Service (QoS) Metrics (Fig. \ref{fig:overview_fig}[1A])} \label{sec:qos_metrics}
The monitoring tools collect a large set of service metrics $\mathcal{M}_{tot}$ for a system. However, many such metrics recorded by these tools are often never used by the SREs \cite{MetricsT9}. Also, storage and handling  of metrics data is non-scalable and gets expensive over time. Consequently, we derived a condensed subset of metrics, denoted as $\mathcal{M}$. In our specific scenario, we filtered down the number of metrics from $\sim$2000 in $\mathcal{M}_{tot}$ \cite{Metrics_aws}  to 42 using a step-wise procedure. 

We employed established techniques \cite{pudjihartono2022review} for feature selection process. Firstly, features were filtered using correlation analysis and rank coefficient tests. Then, time series features that were constant throughout the time series or exhibited low variance were omitted due to their limited informational value. To refine our feature set further, we incorporated domain-specific knowledge: retaining only those metrics that either trigger alerts or have been emphasized in previous outage analysis reports that are generated post-identification and mitigation of outages by engineers. This process yielded a focused feature set well-suited for effective service monitoring and analysis.

However, only a fraction of $\mathcal{M}$ directly reflects the service quality as perceived by the customer, for example, latency of a service, number of service failure errors, resource availability, etc. These metrics, known as \textit{Quality of Service (QoS)} metrics $\mathcal{M}_{QoS}$ or the golden metrics~\cite{googlesrebook}, are used by the SRE to define outages. These metrics are crucial to monitor because cloud service providers face revenue loss if QoS is not met due to violations of Service Level Agreements (SLAs). Based on the alert severity used by the SRE team and the SLA definitions, we select five golden metrics comprising of 
 (i) Workload, (ii) CPU Utilization, (iii) Memory Utilization (iv) Latency, and (v) Errors. These metrics are often used for system monitoring in industries and have been utilized in prior works~\cite{li2022causal, chakraborty2023causil}. The golden signals can often refer to different metrics based on service components. For example, the latency metric refer to disk I/O latency for storage service, web transaction time for web services, query latency for databases, etc. 

\ourmethod{} uses the entire set of metrics $\mathcal{M}$, to forecast the likelihood that $\mathcal{M}_{QoS}$ metrics will surpass a threshold in the future. We do not specifically forecast the likelihood of metric values of $\mathcal{M} \setminus \mathcal{M}_{QoS}$ crossing the threshold since these capture small issues which propagates within the system and gets manifested into the QoS metrics. Also, QoS metrics capture the user impact directly. It should be noted that our choice of $\mathcal{M}_{QoS}$ is based on system domain knowledge which we gathered from the inputs from reliability engineers on the most important metrics that define an outage. Nonetheless, our approach will work in the same way for a different set of $\mathcal{M}_{QoS}$ metrics. 

% \textcolor{red}{
% Why QoS metrics used for outage?
% Why not the other 42 metrics that are monitoring the system health if they can actually act as leading indicators?}

% \textcolor{red}{Logic:\\
% 1.QoS metrics capture SLAs, user impact, they are used for calling out outages, and deteriorate when the system is under heavy stress (extreme conditions).}\\
% \textcolor{red}{2.The other 42 metrics capture a lot of small issues occurring in the system and can be misleading. These 42 metrics are good indicator of issues that can happen when somethings propagate across them and manifest in QoS metrics. Such relationships are captured by our approch.}\\
% \textcolor{red}{3.Monitoring all can can lead to fatigue without any significant improvement in detection.}

\subsubsection{Pre-processing (Fig. \ref{fig:overview_fig}[1B])} \label{sec:metric_preprocess}
After the selection of metrics $\mathcal{M}$, we handle the missing values differently for different category of metrics. For some metrics, a missing value might indicate a null value, which can be replaced with a zero. For other metrics, the rows containing missing values may be dropped. For example, if there are missing values in a metric that defines an error, these can be replaced with zeroes, as this indicates that there were no errors in the service. However, if there are missing values in utilization-based metrics, it may be necessary to drop those rows, as the missing values could be due to a fault with the monitoring system. Once the missing values are handled, each metric $m^i$ is normalized using Equation \ref{eq:normalize}.

\begin{equation}
    m^i = \frac{m^i - min(m^i)}{max(m^i) - min(m^i) + \epsilon}
    \label{eq:normalize}
\end{equation}

Following this, we create a time series of $\mathcal{M}$ metrics with a rolling-window of size $w$. That is, for each time instant $t$ and metric $m^i \in \mathcal{M}$, we create a time series $m^i_w = \{m^i_{t-w}, \ldots, m^i_t\}$, where $m^i_t$ refers to the value of the metric $m^i$ at $t^{th}$ time instant. We thus create $X = \{m^1_w, m^2_w, \ldots, m^{|\mathcal{M}|}_w\}$, which forms a sequence of metric values that can be used as an input to our encoder model.

\subsubsection{Label Generation (Fig. \ref{fig:overview_fig}[1C])} \label{sec:label-generate}
In real-world production services, outages are rare due to the robust deployment architecture and constant monitoring system in place. SREs often intervene to prevent the full-scale outages resulting in a rarity of such events. However, these potential issues when interventions are performed can still be considered as extreme situations (see \S \ref{sec: Background and Problem Definition}), which will allows us to better understand the system's behaviour and predict critical issues in advance. Thus, instead of having the time periods when an outage was actually declared as the ground truth, we modify our definition of \textit{labels} to the time periods of extreme events. Such modifications facilitate us in forecasting the distribution of the relevant metrics. However, the challenge of labelling the data during these extreme events remain. To address this issue, we perform the following algorithmic steps that incorporates domain knowledge to generate proxy labels for outages or extreme situations.

% \begin{enumerate}
% \item From the set of metrics $\mathcal{M}$, select all time window of length $w$ where the metric values \textcolor{red}{which metric values... any one or all?} is greater than a threshold $T$ for a total of $\alpha$ duration.
% \item Filter the time windows from step (1) by checking for the presence of severity alerts that were triggered during those periods. That is, whether those time windows actually resulted in the firing of some high severity alerts.
% \item \textcolor{red}{We cluster the filtered set by metric, alert, and service, and have them reviewed by a human expert (from the SRE team) to obtain the final set of labels}
% \end{enumerate}

\begin{enumerate}

\item Take ${w'}$ minutes windows for each of the metrics ${m}_{i}$ from the set $\mathcal{M}_{QoS}$.
\item Select those windows where the value of ${m}_{i}$ crosses a percentile threshold $\mathcal{T}$ for  at least $\alpha$ fraction of $w'$ window.
\item Filter the previously obtained time windows by keeping only those where at least $\textit{k}$ alerts were fired in the system.
\end{enumerate}

These chosen time windows serve as proxy labels for extreme events. Here, we take $w'$ as 10 minutes, $\mathcal{T}$ as 95, $\alpha$ as $0.5$, $\textit{k}$ as 1.

These steps indirectly incorporate domain knowledge to accurately generate labels for outages and extreme situations using alerts defined by SRE. This process not only allows us to create a denser labelling of extreme events, which can aid in the prediction of potential outages or situations that could have escalated to an outage in advance, but it also includes some less severe cases, which can aid in model training recall. The proxy labels serve as positive training samples for the model.

\subsection{Distribution Forecasting} \label{sec:dist-forecasting}
Through this module, \ourmethod{} aims to learn how the QoS metrics will behave in a future time to forecast the probability of an outage. We outline the component details below.
% More concretely, the Metric-Encoder in Figure \ref{fig:overview_fig} captures latent information in the metric variations, which is then used to train the MDN network to learn the distribution of the metric values of QoS metrics. With a distribution learnt for each of the QoS metrics, we find the probability of an outage by computing the probability of the metric value crossing a certain threshold. 

\subsubsection{Metric-Encoder (Fig. \ref{fig:overview_fig}[2A])} \label{sec:metric-encoder}
Before we can learn the distribution of QoS metrics, we must encode the past behaviour of the service metrics which captures the system state as a latent vector representation. Metric-Encoder extracts information via ML technique to encode spatial as well as temporal relation \cite{shao2020incorporating} between the metrics. Spatial correlation captures how each behaviour of metric $m^i \in \mathcal{M}$ affects the QoS metrics $\mathcal{M}_{QoS}$, while the temporal dependence captures the time series trend in $X$. Though both statistical and ML-based techniques have been studied in this regard~\cite{chatfield2000time, hyndman2008automatic, lim2021time, torres2021deep}, it has been shown that ML-based models, and especially Recurrent Neural Network (RNN) models outperform conventional methods~\cite{siami_arima} in encoding a time series due to their ability to capture sequential dependencies and temporal patterns in data. Several RNN-based models like Long Short-Term Memory (LSTM)~\cite{hochreiter1997long} or Bidirectional LSTM (BiLSTM)~\cite{graves2005framewise, 650093} models can be used for our purpose. Based on our experimental results with various RNN architectures (see \S\ref{sec:metric-encoder-exp}), we choose BiLSTM as the metric encoder model. 

% LSTM maintains a cell state $C_t$ and gating mechanisms to control the amount of information to be encoded from the input time series. The hidden state $h_t$, a multi-dimensional vector, contains the entire encoding of the time series input data. BiLSTMs are an extension of the LSTM models in which two LSTMs are applied to the input data, one in the forward direction of the input sequence, while the second in a reverse direction. The hidden state of a BiLSTM encodes the characteristics of the input sequence and captures information from forward as well as reverse passes. The Metric-Encoder takes $X = \{m^1_w, m^2_w, \ldots, m^{|\mathcal{M}|}_w\}$, which is a $w$ duration windowed sequence of all metric values in $\mathcal{M}$ as input, and outputs a multi-dimensional vector representation ($h$) capturing the temporal and spatial relationship of the metrics.

LSTM uses gating mechanism to control information encoding, while the hidden state ($h_t$), a multi-dimensional vector, maintains the encoding of the input time series. BiLSTMs extend LSTMs by applying two LSTMs, one forward and one backward, to input data to capture information from both directions. The Metric-Encoder takes $X$ as input and outputs a vector representation ($h$) capturing the temporal and spatial relationship of the metrics.

\subsubsection{Multi-Task Learning} \label{sec:multi-task}
We propose a multi-task learning~\cite{ruder2017overview,caruana1997multitask} problem, where one task is to learn the distribution of each QoS metric $y \in \mathcal{M}_{QoS}$ from the \textit{Metric-Encoder} output, while the other task classifies the \textit{Metric-Encoder} output as an outage or not. We now describe each of the task in detail.

\textbf{Task 1: Distribution Learning (Fig. \ref{fig:overview_fig}[2B]).}
The first task aims to learn a parametric distribution governing the QoS metrics conditional on the encoded system state representation. More precisely, given a time series of metrics $X$ which was encoded to form a vector $h$, we wish to estimate the probability of a metric $y \in \mathcal{M}_{QoS}$ given $X$, $p(y | X)$. Learning a distribution is essentially learning the parameters governing it.
In general, the metric $y$ is often assumed to follow a normal distribution $\mathcal{N}(y; \mu, \sigma)$, since we observe limited data points in the tail of the distribution. However, in a real production system, a normal distribution might underfit the actual data distribution. Often, we don't necessarily have simple normal distributions. To overcome this limitation, we estimate the distribution of the QoS metric via a mixture of normal distributions with $C$ mixture components, where the probability distribution of a metric $y$ given $X$ is of the form:
\begin{equation}
    p(y|X) = \sum_{c=1}^C \alpha_{c,y}(X) \mathcal{N}(y|\mu_{c,y}(X), \sigma_{c,y}(X)),
\end{equation}
where $c$ denotes the index of the corresponding mixture component, and $\alpha_{c,y}$ is the mixture proportion  representing the probability that $y$ belongs to the $c^{th}$ mixture component $\mathcal{N}(y | \mu_{c,y}, \sigma_{c,y})$.

It is well known in the literature that a mixture of Gaussian/normal distributions is capable of modelling any arbitrary probability distribution with correct choice of $C, \alpha_c, \mu_c$ and $\sigma_c$~\cite{Reynolds2009}. We aim to estimate a mixture distribution for each QoS metric $y$ via a separate Mixture Density Network (MDN), that comprises a feed-forward neural network to learn the mixture parameters $\mu_y, \sigma_y$ and the mixing coefficient $\alpha_y$. We have chosen $C=3$ after experimental analysis (see Fig. \ref{fig:add-exp1}), and hence each MDN has 3 values of $\alpha_y, \mu_y$ and $\sigma_y$. The network is learnt through minimizing the negative log-likelihood loss of obtaining the ground truth metric value of $y$ given the mixture distribution, averaged over all metrics $\mathcal{M}_{QoS}$. Formally,

\begin{equation}
    \operatorname*{arg\,min}_\theta l(\theta) = -\frac{1}{|\mathbb{R}|}\operatorname*{\sum}_{X,y \in \mathbb{R}} \log p(y|X)
    \label{eq:nll}
\end{equation}

Here $\mathbb{R}$ corresponds to the realm of possibilities. MDN hence learns the parameters of the distribution of QoS metrics, which can then be further used to compute the probability of $y$ crossing a certain threshold to predict outages. 

\begin{figure}[h]
    \centering
    \includegraphics[width=0.25\textwidth]{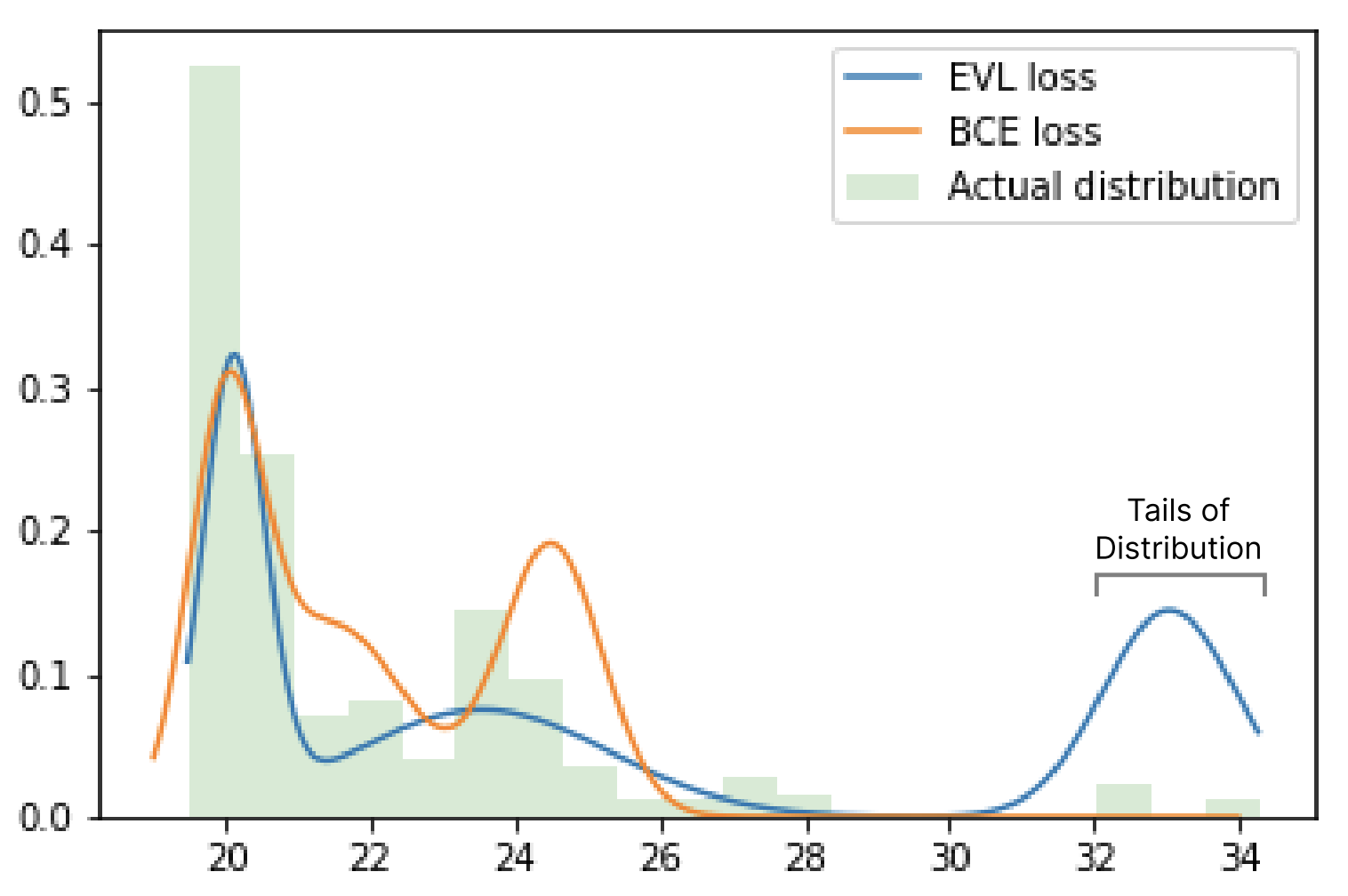}
    \caption{Using Extreme Value Loss in the classifier over BCE can aid the distribution learner to learn a better distribution at the tail.}
    \label{fig:bce-vs-evl}
    \vspace{-0.3cm}
\end{figure}

\textbf{Task 2: Outage Classification (Fig. \ref{fig:overview_fig}[2C]).}
We have observed through experiments (see \S\ref{sec:evl-exp}) that the distribution learnt by MDN performs poorly at the tail, where extreme values are generally observed and can be used to forecast outages (Figure \ref{fig:bce-vs-evl}). To overcome this limitation, a feed-forward neural network performs outage classification in a multi-task setting, where we predict whether an outage will happen or not from the encoded output from the \textit{Metric-Encoder}. We use the output proxy labels generated in \S\ref{sec:label-generate} as a ground truth. 

This module acts the extreme value regularizer. 
where the intuition is that the synthetically generated proxy labels will act as a regularizer for better learning in the tail of the distribution. Similar to distribution learning, we have separate neural networks for each QoS metric in $\mathcal{M}_{QoS}$.

% \begin{figure}
%     \centering
%     \includegraphics[width=0.4\textwidth]{illustration of evl loss}
%     \caption{Caption}
%     \label{fig:my_label}
% \end{figure}

To classify outages, we have used the Extreme Value Loss (EVL)~\cite{ding2019modeling,coles2001introduction}, which is a modified form of Binary Cross Entropy (BCE) Loss as the loss function. EVL reduces the number of false positives by assigning more weight to the penalty of incorrectly predicting outages. EVL works well with imbalanced data as we have observed through experiments. EVL can be formally defined as, 
% \begin{equation}
%     \mathbb{L}_{BCE} = - \frac{1}{N} \sum_{i=1}^N y_i\times\hat{y_i} + (1-y_i)\times \log(1- \hat{y_i}),
% \end{equation}

\begin{multline}
    \mathbb{L}_{EVL} = - \frac{1}{N} \sum_{i=1}^N \beta_{0}\left [1-\frac{\hat{y_i}}{\delta} \right ]^\delta y_i \log{\hat{y_i}} + \\ \beta_{1}\left [1-\frac{1 - \hat{y_i}}{\delta} \right ]^\delta (1-y_i) \log{(1-\hat{y_i})},
\end{multline}

where $N$ is the size of the batch, $y_i \in \{0,1\}$ is the ground-truth value and $\hat{y_i}$ is the value predicted by our model \ourmethod{}, $\beta_0$ is the proportion of normal events in the batch and $\beta_1$ is the proportion of extreme events in the batch.  We use $\delta=2$ in the loss function for the experiments.

\subsection{Training}
Since we want to predict the probability of an outage in advance and reduce the MTTD, the ground truth metric values and the proxy labels should also correspond to a future time $t + \gamma$. Thus, at a time $t$, the Metric-Encoder takes $X$ which is a time series of all metric values from $t-w$ to $t$ as input. The ground truth value for Task 1 is the metric value for each QoS metric $y$ at time $t+\gamma$ and while for Task 2, we use the proxy label (see \S\ref{sec:label-generate}) computed from the QoS metric values at $t+\gamma$. We train the entire pipeline consisting of the Metric-Encoder, mixture density network and the classifier in an end-to-end fashion.

It should also be noted that with a large $\gamma$, one can aim to predict an outage well in advance, but the distribution followed by the QoS metric will not be accurate. Hence, a careful selection of $\gamma$ is necessary. By our experiments (see Fig. \ref{fig:add-exp2}), we show that $\gamma = 10$ mins works the best for our purpose.

\begin{figure}[t]
    \centering
    \includegraphics[width=\linewidth]{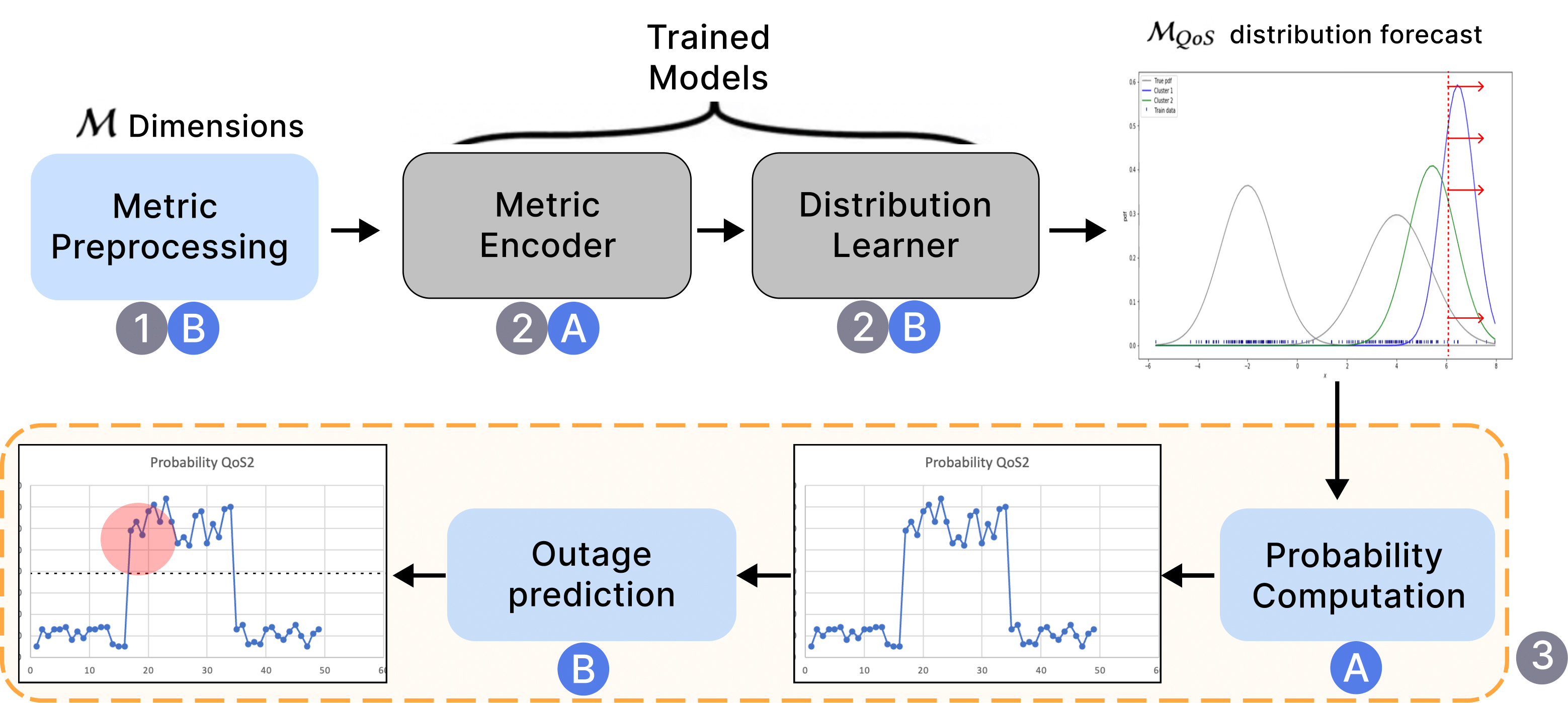}
    \caption{Tasks performed during inference time to predict potential outages from the predicted distribution}
    \label{fig:inference_fig}
    \vspace{-0.4cm}
\end{figure}

\subsection{Inference} \label{sec:inference}
At inference time, we predict and use only the distribution of the QoS metrics, while excluding the classifier from our inference pipeline to predict an outage. The distribution of the QoS metrics provide us with more flexibility and enables us to define outages based on custom thresholds. Moreover, the distribution captures the entire spectrum and specifically the tail metric values. However, the steps to predict an outage from the distribution of the QoS metrics can be summarized as below.

\subsubsection{Probability Computation (Fig. \ref{fig:inference_fig}[3A])}
We first compute the probability of an outage occurring by computing the probability that the value of the QoS metric crosses a pre-defined threshold. These thresholds are generally defined by the SLAs that have been agreed with a particular customer. As an example, an agreement of achieving a service latency of at most $\rho$ milli-seconds for 99\% of the times might have been signed with the cloud service provider, and can be termed as an SLA. Hence, in this case, the threshold is 99\%. Formally, the probability of a QoS metric value $y$ crossing a threshold $\mathcal{T}$, and hence the probability of an outage occurring can be defined as
\begin{equation}
    Prob(Outage) = \sum_{c=1}^C \alpha_c [\mathcal{N}(y|\mu_c, \sigma_c) > \mathcal{T}]
\label{eq:outage-prob}
\end{equation}

\subsubsection{Outage Prediction (Fig. \ref{fig:inference_fig}[3B])} \label{sec:youden}
From the probability computed above, we use a thresholding technique on $Prob(Outage)$ to predict the outages. We compute the threshold based on Youden's J Index~\cite{youden1950index} on training data. It is a popular thresholding technique for imbalanced data (extreme events are very few as compared to the \textit{usual} metric values), which uses the Area under the ROC curve (AUC) to compute the threshold. On the training data containing the proxy labels and the corresponding probability of an outage occurring, Youden's J Index tries to compute the threshold such that it increases the precision and recall. We maintain the same threshold for all our evaluations.

% \subsubsection{Change Point Detection (Fig. \ref{fig:inference_fig}[3B])}
% However, using a threshold on the probability value to detect an outage in a production scenario makes no sense, but the change in probability values provides significant information about a potential outage. Hence, we model a change point detection on $Prob(Outage)$ to detect whether an outage is likely to occur or not. We have used the Pelt (Linearly Penalized Segment)~\cite{killick2012optimal} algorithm implemented in the \textit{ruptures} library\footnote{\href{https://centre-borelli.github.io/ruptures-docs/}{https://centre-borelli.github.io/ruptures-docs/}} to locate points where a change is detected in the probability value. \ourmethod{} predicts an outage when the change is detected in the probability value defined in Equation \ref{eq:outage-prob}.

% \subsection{Deployment}
% \begin{enumerate}
%     \item Train
%     \item Retrain
% \end{enumerate}
% \shubham{Comment:
% Why re-training is needed? How to detect when to re-train? How to re-train?
% Change point detection algorithm points at potential issues, which we detect.
% Validation done by SRE.
% Re-training decided on the basis of it, as system state alters at change point. Retraining is needed.}

%-------------------------------------------------------------------------------

%-------------------------------------------------------------------------------

% \textbf{\shubham{@sarthak We really need to explain: normal situations -> extreme situations -> potential outages -> outages connection, scaling details in metrics preprocessing? (minmax scaler)}}

\section{Implementation Setup} \label{sec:implement}
In this section, we outline the experimental process and the setup we followed. We have implemented \ourmethod{} in \textit{python} and used \textit{tensorflow}\footnote{\href{https://www.tensorflow.org/}{https://www.tensorflow.org/}}~\cite{abadi2016tensorflow}, a standard open-source library to implement the ML models. We have run \ourmethod{} on a system having Intel Xeon E5-2686 v4 2.3GHz CPU with 8 cores.

\textbf{Source of Data:} The data is sourced from a prominent SaaS enterprise offering extensive software and digital services. It leverages Amazon Web Services (AWS) and Microsoft Azure for cloud provisions. The software infrastructure covers diverse domains including programming languages, databases, AWS, Azure, Docker, Kubernetes, Jenkins, and more.

\textbf{Dataset:}
We collect the dataset for evaluating \ourmethod{} from a real-world service hosted by a large cloud-based service provider. The metrics data was obtained through a message queue pipeline deployed on the monitoring system of the service. We have collected a total of 3 months of metrics and outage data from the monitoring system for training and testing purposes where data from the last 3 weeks were used for testing. We collected $\sim$2000 metrics, which was reduced to 42 as discussed in \S\ref{sec:qos_metrics}.
Outages have a widespread impact within the enterprise affecting multiple services. 
Since there were no outages observed during the period of the training data while one outage was observed during the period of test data, we generated time periods when the extreme situations occurred (see \S\ref{sec:label-generate}). It amounted to around 5-7\% of the total training data, thus exhibiting a skewed label imbalance.

\textbf{Model Hyperparameters:}
The implementation details for the ML models used in \ourmethod{} (\S\ref{sec:overview}) are outlined as follows. The BiLSTM model in the Metric Encoder has 128 hidden units ($h=128$), followed by a dropout layer with $p=0.2$. Regularization techniques were used while training the model to prevent overfitting. The feed-forward Mixture Density Network (MDN) which models the distribution parameters of QoS metrics has two hidden layers with 200 neurons each, with ReLU~\cite{glorot2011deep} activation function in the hidden layers. The neuron outputting the mixing factor of components ($\alpha$) use a softmax function. The classifier feed-forward network has one hidden layer with 20 neurons with ReLU activation, while the output layer use the sigmoid function. We use a learning rate of 0.001 with the Adam optimizer for training.

\textbf{Baselines:}
The baselines for evaluation are chosen following an approach similar to the work presented in \cite{lu2020making}. We leverage some of the fundamental classification and regression techniques for outage prediction. It includes Naive Bayes classifier, random forests and gradient boosted decision trees. Naive bayes is a probabilistic machine learning model while the other two are ensemble methods that are constructed using a multitude of individual trees. We implement these baselines to use them as a proxy for prior learning based outage prediction models. We also use a BiLSTM classifier as a baseline, which uses only classifier network on the encoded BiLSTM representation to predict outages.

\textbf{Evaluation Metrics:}
To evaluate the effectiveness of various approaches, we use AUC-PR and F1 score. AUC-PR calculates the area under the precision-recall (PR) curve and is commonly used for heavily imbalanced datasets~\cite{sofaer2019area} where we are optimizing for the positive class (outage being detected) only. AUC-PR is computed using the probability of an outage occurring or not (from Equation \ref{eq:outage-prob}).
Also, based on the probability values, we use the procedure in \S\ref{sec:youden} on training data to compute a threshold for detecting outages, which we use to compute the F1 score in test data. 

\textbf{Other Hyperparameters:}
For all the experiments, we choose a window\footnote{According to our empirical study, over 60\% issues are triggered within 1 hour after the impact start time.} $w = 60$ mins to create a windowed time series of metric data. On the contrary, we vary the prediction look-ahead $\gamma$ from 5 mins to 30 mins. In \S\ref{sec:ablation}, we experimentally show the optimal value of $\gamma$. Unless specified, we maintain threshold $\mathcal{T}=95\%$ (Eq. \ref{eq:outage-prob}) for all the experiments.

%-------------------------------------------------------------------------------

%-------------------------------------------------------------------------------
\section{Evaluation} \label{sec:results}
In this section, we present the experimental results and aim to address the following research questions:
\begin{itemize}
\item \textbf{RQ1:} How do our design decisions align with the ablation studies performed?
\item \textbf{RQ2:} How does our approach compare to the established baselines?
\item \textbf{RQ3:} How does our approach perform in a real-world cloud deployment scenario?
\end{itemize}

\subsection{Design Choices (RQ1)}
\subsubsection{Metric Encoder Model} \label{sec:metric-encoder-exp}
In \S\ref{sec:metric-encoder}, we claim to use Bidirectional LSTM (BiLSTM) as the model for metric encoder. In this subsection, we discuss the experiments conducted to determine the optimal architecture and the rationale behind using BiLSTM. We conducted experiments using four different types of RNNs: LSTM~\cite{hochreiter1997long}, BiLSTM~\cite{graves2005framewise}, Stacked LSTM~\cite{cui2018deep}, and Stacked BiLSTM~\cite{cui2018deep}. The encoded representation was then used to forecast the distribution in a multi-task setting with EVL in the classifier network. The performance of each architecture was evaluated using AUC-PR metric. We have experimented with varying values of $\gamma \in \{5, 10, 15, 30\}$ min. The results are presented in Table \ref{tab:exp1}.

\begin{table}[h!]
\centering
\caption{Design choice: Comparison of different RNN architectures over different prediction windows in terms of AUC.}
\vspace{-3mm}
\begin{tabular}{c|c|c|c|c}
\toprule
\multirow{2}{*}{Model} &  \multicolumn{4}{c}{Prediction Look-Ahead ($\gamma$)} \\
\cmidrule(lr){2-5}
 & 5 mins & 10 mins & 15 mins & 30 mins \\
\midrule
LSTM & 0.950 & 0.950 & 0.950 & 0.948    \\
\midrule
BiLSTM & \textbf{0.974} & \textbf{0.977} & \textbf{0.968} & \textbf{0.959}    \\
\midrule
Stacked LSTM & 0.961 & 0.944 & 0.925 & 0.914    \\
\midrule
Stacked BiLSTM & 0.956 & 0.938 & 0.933 & 0.918    \\
\bottomrule
\end{tabular}%

\label{tab:exp1}
\end{table}

We see that BiLSTM encoder performs the best in our case for all values of $\gamma$. BiLSTM can track a time series in the forward as well as the backward direction. Thus, it can help to encode the overall variation in performance metrics as well as retain recent trends, which makes BiLSTM an ideal choice for encoding the information in the metric time series.

\subsubsection{Multi-Task Learning}
Table \ref{tab:exp2} illustrates that incorporating the proposed multi-task learning approach improves performance compared to using only a single task: classification network or MDN. We used BiLSTM as the Metric Encoder. We evaluate the different schemes using the AUC-PR metric. For the classifier network (individually as well as when evaluated in a multi-task setting), we employed the EVL loss. In a similar setting as of the above, we perform the experiments with $\gamma \in \{5, 10, 15, 30\}$ min.

\begin{table}[h!]
\centering
\caption{Design choice: Comparison of different model architectures over different prediction windows in terms of AUC. Here, MTL refers to the Multi-task learning proposed model.}
\vspace{-3mm}
\begin{tabular}{c|c|c|c|c}
\toprule
\multirow{2}{*}{Model} &  \multicolumn{4}{c}{Prediction Look-Ahead ($\gamma$)} \\
\cmidrule(lr){2-5}
 & 5 mins & 10 mins & 15 mins & 30 mins \\
\midrule
Classifier & 0.909 & 0.914 & 0.930 & 0.927    \\
\midrule
MDN & 0.967 & 0.960 & 0.956 & 0.951  \\
\midrule
MTL & \textbf{0.981} & \textbf{0.982} & \textbf{0.977} & \textbf{0.975}    \\
\bottomrule
\end{tabular}%
\label{tab:exp2}
\end{table}

We observe that when the BiLSTM encoded representation was used to learn a distribution of the QoS metrics in a multi-task setting (learning the distribution and classifying the time periods of extreme values), it performed better than when the tasks were performed individually. This corroborates our design choice of using a multi-task learning in the distribution forecasting module.

\subsubsection{Classifier Network Loss} \label{sec:evl-exp}
Additionally, we perform further experiments to show the performance enhancement of using EVL loss over Binary Cross-Entropy (BCE) loss for the classifier network in \S\ref{sec:multi-task}. The results, as shown in Table \ref{tab:exp3}, indicate that the use of EVL in conjunction with multi-task learning improves performance in predicting extreme events as compared to solely using BCE loss. The metric used to evaluate the performance of the models is the F1-score, and the results demonstrate that EVL outperforms BCE.

\begin{table}[h!]
\centering
\caption{Design choice: Comparison of BCE and EVL loss over different prediction windows in terms of F1 score.}
\vspace{-3mm}
\resizebox{\columnwidth}{!}{%
\begin{tabular}{c|c|c|c|c}
\toprule
\multirow{2}{*}{Model} &  \multicolumn{4}{c}{Prediction Look-Ahead ($\gamma$)} \\
\cmidrule(lr){2-5}
 & 5 mins & 10 mins & 15 mins & 30 mins \\
\midrule
\ourmethod{}(with BCE) & 0.980 & 0.980 & 0.971  &  0.946   \\
\midrule
\ourmethod{}(with EVL) & \textbf{0.987} & \textbf{0.984}  & \textbf{0.974}  &  \textbf{0.954}  \\
\bottomrule
\end{tabular}%
}
\vspace{-3mm}
\label{tab:exp3}
\end{table}

\subsubsection{Ablations} \label{sec:ablation}
\begin{figure}[t]
   \centering
     \begin{subfigure}[b]{0.49\columnwidth}
         \centering
         \includegraphics[width=\columnwidth]{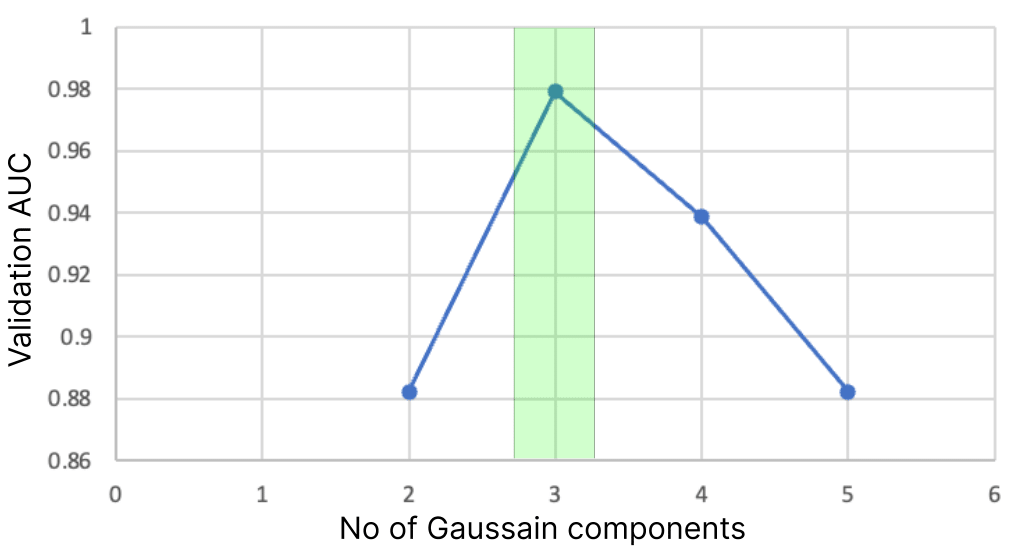}
         \caption{}
         \label{fig:add-exp1}
     \end{subfigure}
     \begin{subfigure}[b]{0.49\columnwidth}
         \centering
         \includegraphics[width=\columnwidth]{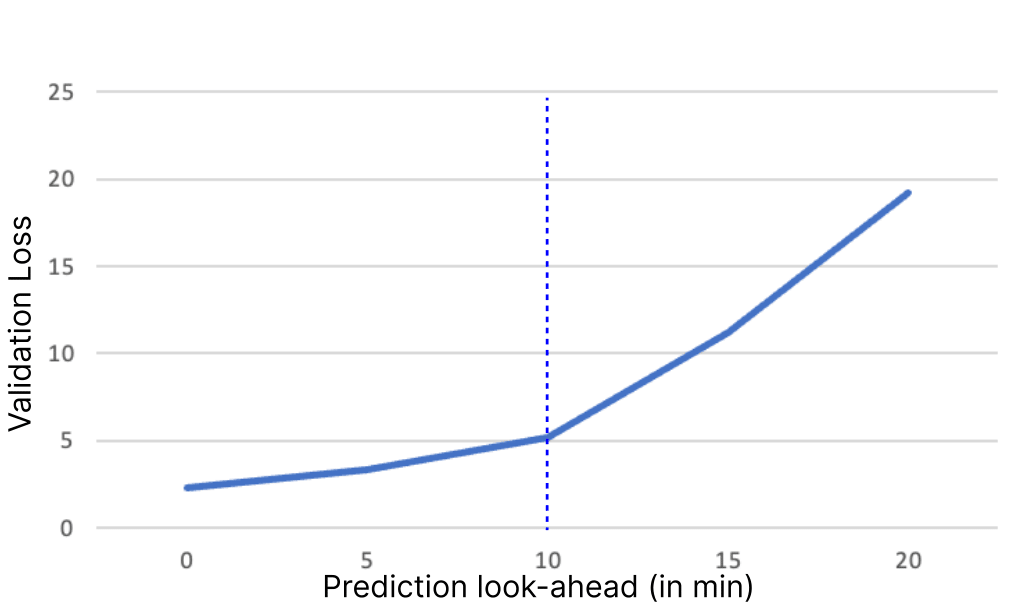}
         \caption{}
         \label{fig:add-exp2}
     \end{subfigure}
    \caption{(a) Model performance vs number of Gaussian mixture components $C$ to predict by MDN; (b) Loss of the MDN (Eq. \ref{eq:nll}) vs Prediction look-ahead $\gamma$}
    \label{fig:add-exp}
    \vspace{-0.4cm}
\end{figure}

We perform further ablation studies to prove our parameter choices for \ourmethod{}. We first illustrate through Figure \ref{fig:add-exp1} that predicting the distribution of QoS metrics using a mixture Gaussian distribution with 3 components performs the best for predicting the outages. We see in the figure that with more or less components, there is a drop in the overall performance. We also conducted an analysis to determine the optimal prediction look-ahead $\gamma$. With large look-ahead $\gamma$, we can forecast the outage well in advance. It however suffers in accuracy of the prediction probability since the inherent trend in the metric changes. Thus, there is a trade-off between $\gamma$ and accuracy metric. Through our experiments, we found that a look-ahead of 10 minutes resulted in the most satisfactory performance, as the validation loss showed negligible increase before reaching a sudden jump beyond this point. Thus, \ourmethod{} can forecast an outage and reduce the MTTD by at least 10 mins than the current approaches (\S\ref{sec:RQ3}).

% \textbf{Answer to RQ1:}
% Our experiments have shown that the Bi-LSTM architecture performs the best as the metric encoder in our system, as seen in Table 1. Additionally, the use of multi-task learning with the EVL loss function (MTL+EVL) improves performance in predicting extreme events, as shown in Table 2 and Table 3. Our analysis of the number of Gaussian components and prediction lag also yielded valuable data specific insights. Over all, the results of these experiments support our hypothesis that the proposed system effectively predicts extreme events with high accuracy.

\subsection{Baseline Comparison (RQ2)}
With our design choices fixed, that is, using a BiLSTM for encoding the metrics and then forecasting the distribution of the QoS metrics in a multi-task learning setting, we compare \ourmethod{} with several baselines as described in \S\ref{sec:implement}. We use AUC-PR to compare the performance and tabulate the results in Table \ref{tab:exp4}. Similar to the previous evaluations, we experiment with multiple values of $\gamma \in \{5, 10, 15, 30\}$ min. The results demonstrate that our proposed approach of forecasting the distribution outperforms all other techniques, including traditional methods, by a significant margin. It has been shown to be a highly effective approach for predicting outages through QoS metrics

One key advantage of \ourmethod{} is its ability to predict the probability of an outage occurring based on any threshold $\mathcal{T}$ (see Equation \ref{eq:outage-prob}) since we are forecasting the distribution as opposed to just learning a classifier with the ground truth proxy labels. On the contrary, traditional methods are limited to predicting outages to a specific threshold (similar to the threshold used for  creating the ground-truth labels in training data). As discussed in \S\ref{sec:label-generate}, we create proxy labels based on the threshold of 95\%, i.e., if the metric value crosses the 95 percentile mark, it is considered to be a potential extreme event. Thus, the classifier network was trained using the generated proxy labels as a ground-truth. However, when we evaluate the distribution forecasted by \ourmethod{} based on the probability of the metric value crossing a threshold of $\mathcal{T}=97\%$ and $\mathcal{T}=99\%$, we achieve high F1 scores, as seen in Table \ref{tab:exp5}. $\gamma$ was maintained at 10 minutes. This flexibility in threshold selection is a major advantage of our proposed method and sets it apart from traditional techniques as the model need not be trained again to get the results on a different threshold.

\begin{table}[h!]
\centering
% \vspace{-4mm}
\caption{Performance of different models over different prediction windows in terms of AUC score.}
\vspace{-3mm}
\begin{tabular}{c|c|c|c|c}
\toprule
\multirow{2}{*}{Model} &  \multicolumn{4}{c}{Prediction Look-Ahead ($\gamma$)} \\
\cmidrule(lr){2-5}
 & 5 mins & 10 mins & 15 mins & 30 mins \\
\midrule
Naive Bayes & 0.593 & 0.592 & 0.592 & 0.582    \\
\midrule
Random Forest & 0.873 & 0.868 & 0.867 & 0.824   \\
\midrule
Gradient Boost & 0.870 &  0.854  & 0.828 & 0.822   \\
\midrule
BiLSTM+Classifier & 0.909 & 0.914 & 0.930 & 0.927    \\
\midrule
\ourmethod{} & \textbf{0.981} & \textbf{0.982}  & \textbf{0.977}  &  \textbf{0.975}  \\
\bottomrule
\end{tabular}%
\label{tab:exp4}
\end{table}

\begin{table}[h!]
\centering
\vspace{-4mm}
\caption{Performance of proposed model over different percentile thresholds in terms of F1 score.}
\vspace{-3mm}
\begin{tabular}{c|c|c|c}
\toprule
\multirow{2}{*}{Model} &  \multicolumn{3}{c}{Prediction Thresholds ($\mathcal{T}$)} \\
\cmidrule(lr){2-4}
 & 95\% & 97\% & 99\%\\
\midrule
\ourmethod{} & 0.984 & 0.972 & 0.968\\
\bottomrule
\end{tabular}%
\label{tab:exp5}
% \vspace{-1cm}
\end{table}

\begin{figure*}[t]
    \centering
    \includegraphics[width=\textwidth]{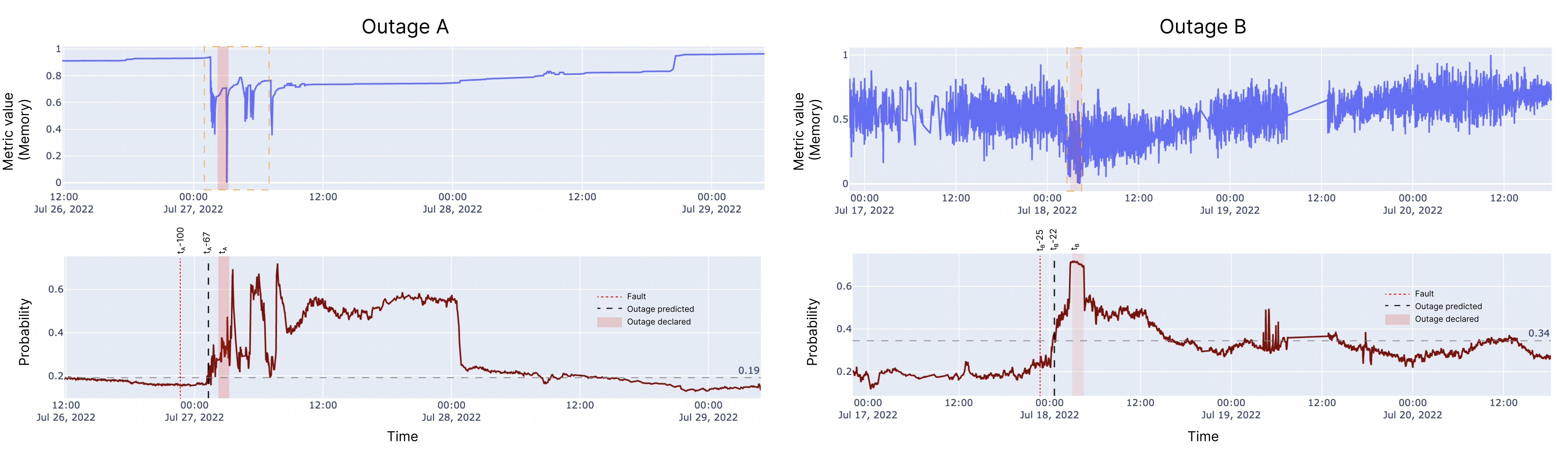}\
    \caption{Analysis of the \ourmethod{}'s performance on two real outages that happened during the deployment period. The upper plots are the metric values which shows deviations (actual value is masked), while the lower plots compute the probability of the metric value exceeding a threshold of $\mathcal{T} = 99\%$. \ourmethod{} was able to correctly predict both the outages in advance (in comparison to the current approach which is indicated by the light red rectangular shaded region).}
    \label{fig:real_data}
    
\end{figure*}

\subsection{Deployment Results (RQ3)} \label{sec:RQ3}
We deployed \ourmethod{} in an enterprise system for 2 months and predicted the probability of impending outages. The overall objective of \ourmethod{} is to predict outages in timely manner, thereby assisting the engineers. From the forecasted distribution, we first predict the probability of a metric value crossing the threshold $\mathcal{T} = 99\%$. We then predict potential outage situations through the thresholding technique as described in \S\ref{sec:youden} on the metric probability that crosses the threshold first. The threshold is generated based on the 9 weeks of training data. We report the precision and recall of the prediction made by \ourmethod{}. We also report the reduction in time to detect outages by the model against the current reactive approach which is used to report an outage.

% We also gather feedback from SRE engineers for any change points detected that were not identified as outages, as not all potential outage situations are reflected in ground truth labels. We use this feedback to calculate the accuracy of predicting potential outages by considering both actual outage labels and potential situations as ground truth.  

% Additionally, to gain a deeper understanding of outages, we further analyze them based on their root cause and detection technique (as outlined in Section \ref{?}).

In this deployment, we implemented a continuous re-training strategy for \ourmethod{}, updating the model after every outage detected with full data for up to two days after the outage ended. This approach was taken to ensure that the system state changes during an outage are reflected in the updated model. Our strategy balances effective model retraining with efficiency. It's worth noting that our focus here does not encompass a robust retraining strategy \cite{wu2020deltagrad} targeting drift issues. The potential for addressing data distribution changes, arising from factors such as the implementation of new business functionalities, could influence outage detection. However, this aspect falls outside the purview of our current work.

We present a case study for multiple outages that were flagged by the engineers during the deployment period and how \ourmethod{} performed in forecasting them. During the deployment period, a total of 4 outages (\texttt{Outage A, B, C, D}) took place, out of which \texttt{Outage A, B} and \texttt{C} manifested through the QoS metrics in the cloud service. \ourmethod{} was able to accurately predict all these three outages and reduced the mean time to detection as compared to the current approach followed by the engineers. However, \texttt{Outage D} was not evident through the QoS metrics and hence it was not predicted. We report the precision, recall and the reduction in MTTD for \texttt{Outage A, B} and \texttt{C} in Table \ref{tab:real-world}. For each outage, we consider data from a day before and 2 days after to report the precision. Precision here refers to the number of outages correctly predicted over the number of times the probability value was above the threshold for a sustained period (15 mins). Recall refers to the number of outages predicted correctly over the total outages that could have been predicted, which is 3. Finally, MTTD reduction for each outage is reported as a percentage of (C - time of prediction)/(C-B) (see Figure \ref{fig:introduction-figure} for the notations of B, C). We observe that \ourmethod{} outperforms other baselines in terms of precision and recall. Recall is 100\%, while precision is 30-40\%. We also observe a large reduction in MTTD\footnote{MTTD improvement: There is a decrease of tens of minutes, particularly notable given the conventional MTTD is also in the range of a few tens of minutes.} for \ourmethod{} (`-' implies outage was not predicted). When EVL is used with \ourmethod{}, precision improves.

\begin{table}[h!]
\centering
\caption{Results for outages predicted by different models using QoS metrics}
\vspace{-3mm}
\resizebox{0.9\columnwidth}{!}{%
\begin{tabular}{c|c|c|c|c|c}
\toprule
\multirow{2}{*}{Model} & \multirow{2}{*}{Precision} & \multirow{2}{*}{Recall} & \multicolumn{3}{c}{Reduction in MTTD} \\
\cmidrule(lr){4-6}
 &   &   & Outage A  & Outage B  & Outage C   \\
\midrule
Naive Bayes & $1/14$ & $1/3$ & $24\%$ & $-$ & $-$   \\
\midrule
Random Forest &$1/11$ & $1/3$ & $0\%$ & $-$ & $-$   \\
\midrule
Gradient Boost & $2/12$ & $2/3$ & $0\%$ & $76\%$ & $-$   \\
\midrule
BiLSTM + Classifier & $2/10$ & $2/3$ & $-$ & $56\%$ & $26\%$   \\
\midrule
BiLSTM + MDN  & $3/10$ & $\textbf{3/3}$ & $43\%$ & $76\%$ & $26\%$    \\
\midrule
\ourmethod{} (BCE) & $\textbf{3/9}$ & $\textbf{3/3}$ & $\textbf{54\%}$ & $76\%$ & $\textbf{27\%}$      \\
\midrule
\ourmethod{} (EVL) & $\textbf{3/8}$ & $\textbf{3/3}$  &  $40\%$ & $\textbf{80\%}$ & $26\%$  \\
\bottomrule
\end{tabular}%
}
\label{tab:real-world}
\end{table}

To provide a deeper understanding of how the system works, we present two case studies of outages (Outage A and Outage B) that were successfully predicted by \ourmethod{}. 

\subsubsection{Outages Predicted} An illustration of these two outages and the performance of \ourmethod{} are presented in Figure \ref{fig:real_data}.
\begin{enumerate}
    \item \texttt{Outage A:} The outage was auto-launched at time $t_A$ by the monitoring systems due to out-of-heap memory issues on several app store nodes in one of the regions. The system \ourmethod{} was able to predict the out-of-heap memory issue correctly and flagged the outage $67$ minutes before the outage was actually launched. The system failed due to a fault in the event consumer queue which got stuck and was not processing since $t_A-100$ minutes in one of the regions.
    % Based on the interviews conducted with the SREs, increased memory size in the application server due to failing nodes was correctly identified as a leading indicator for this issue, with consequences being longer database query time.
    
    \item \texttt{Outage B:} In another outage, alerts regarding high error rates in a service $\mathcal{S}$ were fired. The outage was auto-launched at $t_B$ by monitoring the high error rate. It was found that a faulty update in one of the AWS components had caused the component to fail. That component was being used by $\mathcal{S}$ and therefore after the update, $\mathcal{S}$ was unable to process incoming requests. The issue commenced since $t_B-25$ minutes. \ourmethod{} correctly detected this outage and reduced the MTTD by 22 minutes from the auto-launched approach.
\end{enumerate}

\subsubsection{Outages Not Predicted}
 In addition, we also present a case study of an outage (\texttt{Outage D}) which occurred after a change was implemented that inadvertently tripped an UI feature blocking protocols from making requests. However, such outages due to UI issues are not meant to be manifested in the QoS metrics. As a result, it was not detected by our model. However, this is not a false negative in our case since no changes were observed on the QoS metrics, and hence \ourmethod{} could not have detected the outage. This also highlights the limitations of our proposed method, as it relies on the monitoring metrics to predict outages.

\subsection{Discussion}

From the above results, we see that \ourmethod{} performs better than the other comparable baselines, as well as on real-world deployment setting. However, it can be observed from Table \ref{tab:real-world} that \ourmethod{} reports multiple false positives, which is dependent on the quality of threshold we choose to detect an outage. One can circumvent this issue by having a higher threshold (which might reduce the reduction in MTTD according to Figure \ref{fig:real_data}). However, since we are working with the constraints of data, our training data set did not have any outages and the number of extreme events were very less as compared to the test data. This resulted in Youden's index to compute a threshold lower than 0.5. However, with more data in the training set reflecting the extreme events, thresholding model becomes more proficient to distinguish true positives from the false positives, which results in a higher Youden's index~\cite{sokolova2006beyond}. As and when data arrives, Youden's Index model should be retrained to get an improved threshold. Hence, thresholds can be re-adjusted as well with the predicted distribution, reducing the false positives.

However, in real-world scenarios, the presence of false positive cases is nuanced in this context. Our analysis of model predictions, along with SRE input, identified false positive cases where predicted extreme values didn't lead to outages. Some of these issues were resolved manually or self-corrected. Consequently, the occurrences of false positives are not necessarily negative indicators, as they can capture mild issues that resemble potential disruptions. However, false positives are always undesirable in the workflow of SRE. While SREs can get insights from false positives, their presence is undesirable due to the risk of alert fatigue from frequent, possibly minor, predicted outputs. To tackle this, we provide the benefit of adaptable threshold selection as discussed in \S\ref{sec:inference}. A detailed study on how SREs can distinguish a false positive from a true positive output in their workflow remains an open question.

% In practical scenarios, false positives are always undesirable in the workflow of a Site Reliability Engineer (SRE). However, false negatives are more detrimental to the system than false positives according to the SREs. False negatives would hide the extreme events which can lead to critical outages. Whenever, \ourmethod{} predicts an outage, a low severity alert can be fired, which warns the SREs for a potential extreme event. \sarthak{How can an SRE use the prediction (what if false positive and true positive)? Should we write it from an angle that shows the SRE point of view, which we can get after interviewing them? Or do we write as our own supposition?}

% \shubham{Why not mitigate this from the start by selecting higher thresholds? MTTDs supported by the approach are already relatively high, and the cost in terms of employing SREs to investigate false alerts for high MTTDs can stack from a business standpoint.}

Overall, \ourmethod{} proves to be very helpful for production outage management as it was able to predict the outages well in advance. This could help in reducing the severity and consequently helps with quick mitigation of the outages. We also highlight a limitation of the model, which relies solely on monitoring data and the QoS metrics and cannot predict those outages that are not indicative from these metrics. Nevertheless, we believe that by incorporating additional sources of data, such as log files and change details, we can improve the performance of our model in detecting these types of outages. Overall, our proposed method is a valuable tool for predicting outages using QoS metrics and has the potential to improve system performance and reliability.

%-------------------------------------------------------------------------------

%-------------------------------------------------------------------------------
\section{Conclusion} \label{sec:conclusion}

In this paper, we present a novel approach \ourmethod{} for predicting outages by forecasting the distribution of the relevant QoS metrics. This approach takes a time series input of multiple monitoring metrics representing the current system state within a time window to encode the information present in the time series in a vector representation. It then uses the encoded information to learn the distribution of the relevant metrics using a feed-forward neural network. In addition, \ourmethod{} uses extreme value loss to classify the extreme events in a multi-task manner which helps in learning the distribution of the metrics in the tail. 

At inference time, our model uses the distribution learnt to compute the probability of the metric to cross a certain threshold, and then predicts outages based on a thresholding technique. Our experiments on real-data show the efficacy of our method with an average AUC of 0.98. The applicability and robustness of our approach has been verified by deploying it on an enterprise system.

\textbf{Future Works:} 
Our future works include extending the evaluation duration on production systems to provide insights into long-term performance. Few interesting modifications include automated dynamic threshold selection and providing a confidence bound for our prediction to distinguish between the true positives and the false positive predictions. Furthermore, refining the definition of extreme events could enhance predictive capabilities. Incorporating diverse data sources, such as log files and trace data, may extend the scope of outage detection. We also plan on exploring robust re-training strategies \cite{wu2020deltagrad} to improve model performance in production. Lastly, implementing a feedback loop or introducing human-in-the-loop dynamics may further refine the model's predictive abilities.

\section{Data Availability}

The metrics data used in this research is proprietary and cannot be shared due to confidentiality agreements with the enterprise service provider. However, the model code along with a sample data are made available at \href{https://github.com/skejriwal44/Outage-Watch}{Outage-Watch}\footnote{https://github.com/skejriwal44/Outage-Watch}. The sample data includes the format of the input data required for the model with random metric values. Any real dataset can be pre-processed to the specified format for implementing the approach. We believe that the model code and sample data provided in the paper are sufficient for replication and working with similar datasets. 

\newpage
%-------------------------------------------------------------------------------

\bibliographystyle{ACM-Reference-Format}
\balance
\bibliography{main}

% \appendix

\end{document}